\documentclass[smallextended]{svjour3}     
\smartqed  
\usepackage{graphicx}
\usepackage{color}
\usepackage{amsmath}

\newcommand{\dB}{\ensuremath{\nabla \cdot {\bf B}}}

\journalname{General Relativity and Gravitation}

\begin{document}

\title{Numerical relativity simulations in the era of the Einstein Telescope}

\titlerunning{Numerical relativity and ET}        

\author{Mark Hannam         \and
        Ian Hawke 
}

\institute{Mark Hannam \at
  Physics Department, University College Cork, Cork, Ireland \\
           \and
           Ian Hawke \at
           School of Mathematics, University of Southampton, SO17 1BJ,
           UK
}

\date{Received: date / Accepted: date}

\maketitle

\begin{abstract}
Numerical-relativity (NR) simulations of compact binaries are expected to
be an invaluable tool in gravitational-wave (GW) astronomy. 
The sensitivity of future detectors such as the Einstein Telescope
(ET) will place much higher demands on NR simulations than first- and
second-generation ground-based detectors. We discuss the issues 
facing compact-object simulations over the next decade,
with an emphasis on estimating where the accuracy and parameter space 
coverage will be sufficient for ET and where significant work is needed.
\end{abstract}

\section{Introduction} 

Experiments aimed at the first direct detection of gravitational waves are now
underway with a network of  laser-interferometric gravitational-wave (GW)
detectors~\cite{Abbott:2007kv,Acernese2006,GEOStatus:2006}. 
At the completion of the current science run the LIGO and 
Virgo detectors will be upgraded to \emph{second generation} detectors, with
an order of magnitude improvement in sensitivity, and Advanced LIGO is due to 
go online in 2014. It is hoped that the second-generation detectors will yield
enough observations to begin in earnest the field of GW astronomy; 
see~\cite{Sathyaprakash:2009xs} for an overview of the astrophysics and 
cosmology potential of GW observations.

With the goal of achieving a further factor of ten improvement in sensitivity, 
combined with an extension of the detector bandwidth to the range 1Hz to 10kHz,
a design study for a \emph{third generation} detector has been supported within
the European FP7 framework. If these design goals are met, the Einstein 
Telescope (ET) would be sensitive to a volume of the universe one million 
times larger than current ground-based detectors, and allow a precision 
probe of gravitational-wave sources throughout the universe to
cosmological distances. 

Extracting the most science from ET observations will require 
accurate theoretical knowledge of GW sources --- far more accurate than that
required for first- and second-generation detectors. One 
important aspect of source modelling is the numerical solution of Einstein's 
equations for the inspiral and merger of compact bodies (black holes and
neutron stars) and the numerical simulation of the physics of single compact
bodies (neutron star oscillations, instabilities and collapse, and
core-collapse supernovae amongst others). 

NR simulations have advanced significantly in the last few years,
particularly in describing the last orbits and merger of systems of
two black holes (see
\cite{Pretorius:2005gq,Campanelli:2005dd,Baker:2005vv} for the
original breakthrough results, and \cite{Hannam:2009rd} for an
overview of the status of current simulations focused on GW
detection).  The mode-ling of neutron-star binaries has also made
significant progress from the first simulations of the Tokyo group
(\cite{Shibata:1999wm}) to the current status as recently reviewed
by~\cite{0264-9381-26-11-114004}.

In this paper we will discuss some of the issues that numerical-relativity
simulations face in producing the results needed for ET data analysis
and astrophysics applications. 

For black-hole binary simulations, the 
situation is easy to state: since the total mass of the system provides an
overall scale to the numerical results, the increased bandwidth of ET
does not change the physical nature of the simulations that need to 
be produced; the black-hole binary parameter space remains the same.
However, it does change the accuracy requirement of the simulations, 
and of the model of the GW parameter-space that can be produced from them.

For matter simulations, the situation is quite different. The scales
are set by the physics included in the model. The size of the
parameter space is dominated by the unknown physics which it is hoped
that GW observations will constrain.  In order to make simulations
practical, only the ``essential'' physics will be included in the
model. The determination of ``essential'' is balanced between current
knowledge and expectations of the key physics, and what can
practically be simulated and detected.  Thus the model, resulting
parameter space and accuracy requirements will change with the
increased bandwidth of ET.

In Section~\ref{sec:bbh} the status of vacuum simulations of binary
black holes is discussed, including a discussion of the waveform
accuracy of current and near-future simulations, the sampling of the
parameter space, the construction of template banks and possible
future improvements in numerical techniques. Section~\ref{sec:matter}
discusses the situation with matter simulations, assuming that the
physical model is ``added on'' to the spacetime simulation and GWs
extracted using the same techniques. Section~\ref{sec:discussion}
summarises the likely status of NR over the next decade.

\section{Black-hole binaries}
\label{sec:bbh}

Fully general-relativistic simulations of the dynamics of black-hole binary
systems during their last orbits and merger have been possible since 
2005~\cite{Pretorius:2005gq,Campanelli:2005dd,Baker:2005vv}. 
In these simulations initial data are prescribed for a slice of spacetime
that contains two black holes, usually in orbit, and in many cases with 
initial parameters chosen, or tuned, such that the black holes are following
non-eccentric quasi-circular 
orbits~\cite{Pfeiffer:2007yz,Husa:2007rh,Campanelli:2008nk}. 
The initial data must also satisfy 
four constraint equations in order to be part of a valid solution of Einstein's
equations. The data are advanced to future time slices via a
set of evolution equations; the specific equations depend on the choice of
decomposition of the four-dimensional Einstein equations to 3+1 
(space+time) dimensions, and the coordinates and time-slicing at
successive evolution steps depend on the choice of gauge conditions.
We are in principle free to choose the gauge conditions as we wish 
(as a result of the coordinate invariance of Einstein's equations), but 
in practice are limited to choices that lead to sufficiently stable and
accurate numerical simulations. The issues involved in finding a 
numerically well-posed and stable formulation of Einstein's equations,
a convenient geometrical representation of a black-hole spacetime, 
construction of black-hole-binary initial data, and numerically and 
geometrically well-behaved gauge conditions, are discussed in the
textbook~\cite{Alcubierre2008}, and in the review 
articles~\cite{Cook:2000vr,Pretorius:2007nq,Husa:2008jx}.

The parameter space of inspiraling black-hole binaries consists of ten 
parameters: the total mass and mass ratio of the system (or alternatively 
the individual masses of the two black holes), the spin vector of each black
hole, the eccentricity of the system, and the initial separation of the binary 
(or alternatively the initial phase of the GW signal). Since the total mass
provides an overall scale for the solution, this can be removed from 
the parameter space of necessary numerical waveforms. 
If the numerical late-inspiral-plus-merger waveform is 
combined with an analytic inspiral waveform (for example, calculated
from the post-Newtonian approximation), the initial separation of the 
binary can be made arbitrarily large, and yet one more parameter can be
removed from the parameter space. Assuming that the process of
constructing a ``hybrid'' waveform is sufficiently robust (and this remains
a topic of current research), we are left with a total of eight parameters.  

To date late-inspiral-merger-ringdown simulations of black-hole binaries have 
been performed for systems with mass ratios up to 1:10 (although most
do not extend beyond 1:4), and systems with a variety of spin 
magnitudes and orientations, and for several choices of initial 
eccentricity. A periodically updated summary of simulations that include at 
least ten GW cycles before merger is given in the online version of the
review~\cite{Hannam:2009rd}. Note that although a large number of simulations 
have been performed, these only account for an extremely sparse sampling 
of the full parameter space, and do not yet include any simulations of mass 
ratios beyond $q = m_1/m_2 = 10$, or of spins above $a/m \sim 0.92$. 
Efforts to use numerical-relativity simulations to make study the 
astrophysics of electromagnetic counterparts have also 
begun~\cite{Palenzuela:2009yr,vanMeter:2009gu}.

The issues for black-hole-binary simulations, as they relate to GW astronomy,
are: (1) determining and achieving sufficient accuracy to extract the maximum
physical information from GW observations, which includes producing simulations
that include enough cycles that subsequent analytic-numerical hybrid waveforms
meet these accuracy requirements, (2) producing simulations of a sufficiently
dense sampling of the parameter space, (3) developing methods and codes
that are efficient enough to achieve these goals. 

In distinguishing the requirements on numerical waveforms between ET and
other detectors, for example Advanced LIGO, the key difference is that since 
ET is expected to be roughly an order of magnitude more sensitive than Advanced 
LIGO, the waveform accuracy will also need to be higher to allow the most science
to be extracted from observations. The question for numerical relativity then
is what those accuracy requirements are, and whether they can be
achieved before ET is completed.

\subsection{Waveform accuracy} 
\label{sec:bbhaccuracy}

In the last few years questions of waveform accuracy requirements have begun
to be addressed~\cite{Lindblom:2008cm}, and it seems that current
methods and simulations are sufficiently accurate for the current generation
of ground-based detectors, although a systematic study has been
performed only for the case of equal-mass nonspinning 
binaries~\cite{Hannam:2009hh}. 
In that work it was shown that, within certain reasonable caveats, it will not be 
possible to use GW observations to distinguish between the most accurate
numerical waveforms used in the study unless the signal-to-noise ratio (SNR) is 
above about 25. 

SNRs of that magnitude are expected to be rare from the Initial and Enhanced 
LIGO and Virgo detectors. However, the Einstein Telescope is expected to be 
an order of magnitude more 
sensitive, and SNRs in the range 100-1000 may be typical. We first
consider whether numerical simulations will be able to match the
accuracy requirements of ET by the time of its 
construction.

Waveform accuracy requirements are discussed in some detail in~\cite{Lindblom:2008cm}. 
One simple criterion for waveforms to be sufficiently accurate for all 
parameter estimation purposes is that the \emph{average}
error in the amplitude and phase (appropriately weighted by the power spectrum 
of the detector noise~\cite{Lindblom:2008cm}) satisfy \begin{equation}
\bar{\delta \phi}^2 + \bar{\delta \chi}^2 < \frac{1}{\rho^2},
\end{equation} where $\rho$ is the SNR, and $\bar{\delta \chi}$ is the fractional 
error in the amplitude and $\bar{\delta \phi}$ is the error in the phase. If we
assume that most black-hole-binary signals of interest for ET will have an
SNR of less than 1000, then we require that the average phase and 
fractional amplitude errors be of the order of $10^{-3}$. For the waveforms
described in~\cite{Reisswig:2009us} (as an example of a state-of-the-art
simulation), the maximum phase and amplitude errors
are roughly an order of magnitude larger than the average errors; we
expect that the detector-noise-weighted average errors will behave
similarly. 

As such, the most accurate \emph{numerical} simulations to
date~\cite{Scheel:2008rj} have total phase and amplitude errors of
0.02~radian and 0.3\% respectively (see the table in~\cite{Hannam:2009hh} for
a convenient comparison of errors between codes), and so we 
expect that the average errors are already within the accuracy 
requirements of ET. The simulation that deals most convincingly with
the wave extraction~\cite{Reisswig:2009us} gives average phase and amplitude
errors also within the approximate $10^{-3}$ requirement 
provided here. 

So it appears that state-of-the-art numerical simulations are already at,
or close to, the accuracy requirements for ET science. Assuming that
Moore's Law holds, then the increase in computing power by the time ET
is operational will be $2^{10/1.5} \approx 100$. Current codes typically
employ spatial derivatives that converge to fourth-order or better with 
respect to the spatial resolution, and if we assume fourth-order convergence
in future numerical results, we will see an improvement of two orders of magnitude 
over current results in the next decade. This is a conservative estimate (it
assumes that only the computers, and not the codes, will improve!), and 
serves only to demonstrate that accuracy of numerical methods will not be the
bottleneck in providing waveforms for ET science. 

(It is true that this
simple analysis does not take into account higher modes, for which
comparable accuracy is harder to obtain, but for which high accuracy
will be needed for parameter 
estimation~\cite{Sintes:1999cg,VanDenBroeck:2006ar,Babak:2008bu,Thorpe:2008wh}. 
However, even the
subdominant modes are well-resolved in the most accurate recent
simulations~\cite{Boyle:2008ge,PollneyNRDA}.

As we will see in the next sections, the greater challenge is in sampling the 
black-hole-binary parameter space.

\subsection{Phenomenological modelling of black-hole-binary waveforms}
\label{sec:phenom}

Individual black-hole-binary simulations can only provide a discrete sampling of
the parameter space. The practical construction of template banks for searches,
and the use of theoretical waveforms for parameter estimation and other 
follow-up studies, requires waveforms for \emph{any} values of the black-hole
parameters. As such, the natural approach is to construct an analytic model 
of the waveforms, and to use the numerically generated waveforms as input
to calibrate the free parameters of the model. Such models also aim to 
extend the numerical waveforms to an arbitrarily large number of GW cycles
before merger (or, put another way, extend them to arbitrarily low GW frequency).

To date there have been two approaches to this problem. One is to start with 
an effective-one-body (EOB) model of inspiral-merger-ringdown waveforms (which
in some forms pre-date the existence of full numerical 
waveforms~\cite{Buonanno:1998gg,Buonanno00a,Damour:2000we,Damour:2001tu,Buonanno:2005xu}), 
and to both calibrate unknown EOB coefficients to numerical results, and to 
introduce extra ``flexibility parameters'' that allow yet greater fidelity between 
the EOB model and the full GR results. EOB-based models for nonspinning
waveforms have been pursued by several 
groups~\cite{Buonanno:2007pf,Damour:2007yf,Damour:2007vq,Damour:2008te,Baker:2008mj,Mroue:2008fu,Damour:2009kr,Buonanno:2009qa}, and have reached 
an impressive level of faithfulness to the full GR 
waveforms~\cite{Damour:2008te,Damour:2009kr,Buonanno:2009qa}. 

The second approach has been to construct a purely phenomenological 
ansatz motivated by post-Newtonian (PN) predictions for the inspiral, 
approximate models of the ringdown, and the observed behaviour of the 
signal in the intermediate plunge and merger phase, as seen for
example in~\cite{Buonanno:2006ui}. This model, in which each ingredient
is motivated by known physics, is extremely flexible, and has been 
extended from original work on nonspinning 
binaries~\cite{Ajith:2007qp,Ajith:2007kx,Ajith:2007xh} to binaries
with nonprecessing spins~\cite{Ajith:prep} and has been shown to also be applicable
to neutron-star binaries~\cite{Shibata:2009cn}. This procedure is limited only 
by available numerical simulations, although it remains to be seen how
easily a ``simple'' ansatz can be developed for general precessing-spin
binaries. The nonspinning waveform model is constructed from ten 
mass-ratio-dependent parameters, which can in turn be modelled by 
second-order polynomials, therefore requiring at least three simulations
to define the model. The EOB models, by contrast, can be calibrated by
only one simulation. It remains
to be seen how well this economy of calibration extends to the 
spinning-binary cases.

Both the EOB and pure phenomenological approaches require that the 
waveform model for the GW cycles \emph{before} the beginning of the NR
waveform (i.e., the PN or EOB model) is accurate. This requires comparison 
of NR waveforms with their PN counterparts in the regions where they 
overlap. NR-PN comparisons have been performed for equal-mass
nonspinning binaries~\cite{Baker:2006ha,Hannam:2007ik,Gopakumar:2007vh,Boyle:2007ft},
equal-mass binaries with non-precessing spins~\cite{Hannam:2007wf}, 
eccentric binaries~\cite{Hinder:2008kv}, and one configuration of an unequal-mass
binary with precessing spins~\cite{Campanelli:2008nk}. These results suggest
that PN methods perform well up to the point where NR simulations begin. 
Detailed studies of the accuracy of NR-PN hybrid waveforms based on these
simulations remain to be made, but first studies of the accuracy of equal-mass
hybrids at least suggest that they will fulfil the \emph{detection} requirements
of current ground-based detectors for masses as low as 
10\,$M_\odot$~\cite{Hannam:prep}. Complementary results for studies of pure
PN inspiral waveforms suggest that they in turn will be sufficient for masses 
\emph{up to} 12\,$M_\odot$~\cite{Buonanno:2009zt}. However, the wider
sensitivity band of ET may require that the hybrid waveforms include 
more accurate PN ingredients, or longer numerical simulations. This is a topic
that deserves further study in the coming years.

\subsection{Sampling of the parameter space}
\label{sec:paramspace}

Whichever phenomenological approach turns out to be most practical, it is
clear that simulations covering a fine sampling of the black-hole-binary 
parameter space will be necessary. An exhaustive sampling of the 
parameter space is usually presented as
a monumental computational challenge. Let us consider the scope
of this challenge, and the degree to which it may be met by the time
ET might be commissioned. 

There are eight parameters in the black-hole-binary parameter space.
A recent phenomenological family  of inspiral-merger-ringdown
waveforms~\cite{Ajith:prep} was able to faithfully model binaries with
non-precessing spins using an analytic ansatz in which the coefficients
were expressed as third-order polynomials in the physical parameters
of the model (in this case the mass ratio and a parameter that can be 
approximately considered as the total spin of the binary). 
For simplicity, let us take the relatively optimistic view that third-order
 polynomials in each of the parameters will be sufficient to model the 
 full parameter space, and that we therefore require at least four data points 
 in each direction to robustly constrain our model.
This means that we will require $4^8 \approx 65,000$ simulations. 
This would indeed be a monumental undertaking!

However, this does not take into account symmetries and degeneracies
in the parameter space. For example, the work in~\cite{Ajith:prep} 
successfully parameterizes non-precessing binaries by only the
\emph{total} spin of the binary --- with the consequence that if the 
individual spins aren't necessary to model the waveform, then 
they will also be difficult to distinguish. This reduces the amount of 
information that can be obtained from a detection based on this
waveform model, but also reduces the computational cost of 
producing the model. 

As a second example, the works of \cite{Boyle:2007sz,Boyle:2007ru} 
exploit the 
symmetries of black-hole-binary systems to argue that certain 
quantities in black-hole mergers (the mass, spin and recoil velocity
of the final black hole) can be described by a model that requires
only 26 simulations. Their model deals with only quasi-circular
inspiral (so the parameter space is reduced to seven, not eight,
parameters), and uses a second-order-polynomial ansatz, but 
nonetheless requires far fewer simulations than a naive estimation of 
$3^7 \approx 2200$!

These early attempts to characterize black-hole-binary systems
suggest that a full mapping of the parameter space may be possible
within the next decade, if not the next few years. Before becoming
overly optimistic, however, we should bear in mind that the error 
analyses of many current simulations deal only with the dominant
harmonic (the $\ell = 2,m=2$ mode) of the waveform, and to date
it is only this mode that has been included in analytic/phenomenological
models of black-hole-binary waveforms. Also, two important areas of
the parameter space have not yet been accessed by current simulations. 

One is black holes with high spin. All published long simulations to date
have used black-hole initial data that are conformally flat, an assumption
that simplifies the construction of the data, but limits the spin that can
be modelled. The behaviour of black-hole-binary systems with near-extremal
spins (which may be the most common 
astrophysically~\cite{Volonteri:2004cf,Gammie:2003qi,Shapiro05}), have 
therefore not yet been studied in detail. Proposals have been made to
construct non-conformally-flat initial data for both 
excision-based~\cite{Lovelace:2008tw,Lovelace:2008hd}
and moving-puncture~\cite{Hannam:2006zt} codes, although
only the excision data is sufficiently developed for use in full binary 
simulations. With workable proposals in place for 
high-spin data, it is likely that high-spin binaries will be simulated in 
detail in the next few years, and this region of the parameter space
will be well modelled by the time of ET.

The situation is different for large mass ratios. In current simulations, the
numerical resolution requirements of the simulation are defined by the 
smallest black hole, while the overall scale is set by the total mass. This 
means that the higher the mass ratio ($q = m_1/m_2 \geq 1$), the more 
effective resolution is required, and the more computationally expensive
the simulation. Long, accurate simulations suitable for use in GW applications
have been performed up to $q = 6$~\cite{Baker:2008mj}, and a short simulation 
of $\approx 2.5$ orbits~\cite{Gonzalez:2008bi} exists for $q=10$. 
Since the simulation cost (both in simulation time and memory usage) 
scales \emph{at best} linearly with 
the mass ratio, and more realistically quadratically, long accurate
simulations of mass ratios $q \sim 10$ are at the very limit of current 
methods and resources, and $q \sim 100$ are out of the question. An 
improvement by a factor of 100 in computer power by the time of ET 
suggests that mass ratios $q \sim 30$ may be possible by that time, but
$q \sim 100$ will require new techniques; some possibilities are described
in Section~\ref{sec:improvements}.

It may ultimately not be necessary to simulate systems with $q > 10$ in full
general relativity to accurately model the full parameter space. For example,
the work in~\cite{Ajith:prep} includes simulations up to only $q=3$, but 
models all mass ratios by explicitly incorporating an analytically calculated 
extreme-mass-ratio limit. The effective-one-body (EOB) procedure also 
includes by construction the extreme-mass-ratio
limit~\cite{Buonanno:1998gg}.
 However, in order to
verify that the high mass ratios are indeed being faithfully modelled, it will
be necessary to perform at least a few simulations at very high mass 
ratios. Even if only a few such simulations are required to verify the 
fidelity of a given model (or set of models), new techniques will be
necessary to produce those simulations.

\subsection{Numerical techniques and room for improvement}
\label{sec:improvements}

For codes that employ the moving-puncture method, numerical methods
have not changed significantly since the first simulations in 2005,
but the accuracy and length of the waveforms has improved significantly. 
The first waveforms covered only a few GW cycles before merger, while 
the most recent simulations cover up to 30 cycles before merger. The first
long simulations that allowed comparison with post-Newtonian (PN) 
results had an amplitude uncertainty of around 10\% and an accumulated
phase error of around 1\,radian before merger~\cite{Baker:2006ha}. The most
recent simulations can achieve an amplitude error of 0.3\% and an
accumulated phase error through inspiral, merger and ringdown of 
0.02\,rad~\cite{Boyle:2008ge}. 

Typically mesh-refinement is 
used to resolve both the small-scale features near the black holes,
the medium-scale physics of the GW signal, and the large scales
necessary to push the outer boundary of the computational domain
far from the source. However, improvements in general code efficiency
and the introduction of higher-order spatial finite-differencing 
(first sixth-order in~\cite{Husa:2007hp,Hannam:2007ik}, and later eighth-order 
in~\cite{Lousto:2007rj,Campanelli:2008nk}) have
allowed impressive improvements in both code speed, memory
efficiency, and numerical accuracy. More recent results by two 
groups~\cite{Pazos:2009vb,PollneyNRDA} have demonstrated the 
efficacy of multipatch methods, which allow for a far more efficient
gridding of the wave extraction region, which may lead to yet 
greater accuracy (of the wave extraction) and computational 
efficiency in future simulations. The results of Reisswig {\it et. al.} have
further achieved Cauchy-Characteristic Extraction (CCE), in which the 
waves extracted at some finite spatial distance from the source 
are then evolved to null infinity, which is where GW signals can
be unambiguously defined~\cite{Reisswig:2009us,lrr-2009-3}. 
These results indeed suggest that this method produces results that are 
gauge invariant. 

Simulations with variants of the generalized-harmonic formulation have been
performed by
Pretorius~\cite{Pretorius:2004jg,Pretorius:2005gq,Pretorius:2006tp} 
and by the Caltech-Cornell group~\cite{Lindblom:2005qh,Boyle:2007ft}. 
The latter's SpEC code makes use of a multi-domain spectral method, 
which allows a high degree of accuracy and, like the multi-patch
methods described above, allow high accuracy with minimal 
computational resources in the wave extraction region. 

These codes represent the current state of the art. However, far
greater accuracy and efficiency will be needed to meet the science
goals of ET. Is there room for improvement? 

The answer is certainly Yes. A number of options are yet to be
fully explored. One that has already received some attention is the use
of mixed implicit-explicit (IMEX) time integration to allow far larger
evolution time steps. A first exploration of this approach has already 
been made in the context of scalar fields on a curved 
background~\cite{Lau:2008fb}. Another approach
with a related potential benefit is the use of symplectic 
integrators~\cite{Brown:2005jc,Richter:2008pr,Richter:2009ff,Frauendiener:2008bt}.
In this approach, conserved quantities in the physical system are maintained
by construction, and this in general allows for far larger time steps. The
canonical example is the evolution of binaries in Newtonian gravity, where
dramatic improvements in accuracy are possible. General relativistic systems
are of course far more complex, and there is the added difficulty that 
energy can be dissipated as gravitational radiation, but these issues do not
preclude the use of symplectic methods, and some research on this topic
is already underway. 

A further improvement in both accuracy and memory efficiency could be
achieved by the use of hyperboloidal slices of the spacetime. Here the slices are 
spacelike near the source, but asymptotically null, meaning that they reach null
infinity on a finite numerical domain. Since the GW phase is constant
on a null slice, only a finite number of GW cycles need be resolved between
the source and the outer boundary. This approach would therefore save
computational resources in the wave zone, while at the same time allowing
clean GW extraction at null infinity. Although hyperboloidal initial data for
such systems has already been
proposed~\cite{Ohme:2009gn,Buchman:2009ew}, 
stable simulations are more challenging. Progress has however been made in spherical
symmetry and axisymmetry~\cite{Moncrief:2008ie}, 
and hyperboloidal simulations of black-hole
binaries may be possible in the next few years. 

In the case of moving-puncture simulations, improvements may also be 
possible by better exploiting the natural geometry of black-hole slices in this
approach. The method was developed for data that describe the black holes
as topological wormholes, but it was later found that the slices quickly 
asymptote to cylinders that end at the black-hole throat 
(also called ``trumpets'')~\cite{Hannam:2006vv,Hannam:2008sg,Hannam:2009ib}. 
It may be possible to modify the method to exploit more explicitly this 
trumpet form of the data. In particular, if the original concept of the puncture
approach can be resurrected --- where the singular nature of the data near the 
black holes is described analytically and only small deviations from this form 
(due to the presence of the second black hole) are described numerically --- 
then it may be possible to achieve dramatic improvements in code accuracy and
efficiency.  

These are a number of avenues for improvement that are already being
explored, and have the potential by themselves to make possible large numbers of 
long, accurate simulations possible --- but of course we expect that by 2020 
there will be many other ideas, and it is quite conceivable that by then the 
standard methods of the field will be radically different from those today.

\section{Matter}
\label{sec:matter}

In contrast to simulations of binary black holes, many calculations of
gravitational waves have not used full general relativity (for example
most calculations of core-collapse as reviewed by
e.g.~\cite{Ott:2008wt} and some binary simulations such
as~\cite{Oechslin:2007gn}). This is unsurprising; for scenarios such
as core-collapse supernovae the effect of including the key physics in
the model is more important than a perfect model of gravity. However,
with the recent successes in evolving the spacetime as detailed in
section~\ref{sec:bbh} the use of full general relativity in
calculations of gravitational waves is likely to become standard on
the timescales important for ET.

In addition we note that as the matter terms do not affect the
principle part of the spacetime evolution equations there is only one
potential complication for the spacetime evolution by adding
matter. In vacuum evolutions, away from singularities, all quantities
remain smooth. As the matter quantities may become discontinuous, a
simulation that includes such shocks will have an effect on the
spacetime. It is currently unclear what the impact is in GW
simulations, as the only work to look at this (\cite{Barnes:2004if})
is in $1+1$ dimensions. As no effect has been seen in any current
simulation we will assume that no fundamental problem will affect
matter simulations relevant for ET.

From these basic points we will assume that all current GW simulations
including matter will be able to benefit from the theoretical and
numerical advances for spacetime and gauge evolution and wave
extraction reviewed or proposed in section~\ref{sec:bbh}. Therefore
the bottleneck in computing accurate, realistic GW signals will be in
the physical model of the matter that can practically be evolved, and
the accuracy of the numerical methods used to evolve it.

The parameter space for matter simulations is considerably larger than
for binary black holes. Firstly the underlying physics of the problem
sets the mass scale which cannot be removed. Secondly there are a wide
variety of GW sources; as well as binary systems there are discs and
single star collapse. But most importantly there is the uncertainty in
the physics that the model must include to be realistic (e.g.,
neutrino transport plays a crucial role in core collapse but is much
less important in a binary inspiral), and in the details of that
physics (such as the precise form of the equation of state (EOS)
needed to close the perfect fluid equations).

The approach so far has been to focus on limited areas in the
parameter space to determine which elements of the physical model
affect the qualitative form of the GWs, and if the GW signal is
relatively robust to determine what qualitative effects minor changes
in the parameters make. The key additional parameters in inspirals are
currently believed to be the bulk EOS and the magnetic field; in core
collapse neutrinos and composition are more important than magnetic
fields. Additional effects such as superfluidity, an elastic crust,
diffusion, dissipation, or other non-ideal effects may play a
role. At present even the simplest implementation of these effects is
ongoing research.

The construction of initial data for binary systems is similar to that
for binary black holes and, as reviewed
by~\cite{Cook:2000vr,0264-9381-26-11-114004}, similar issues of the
accuracy of the conformal flatness approximation and how to produce
quasi-equilibrium initial data arise. For single star collapse either
axi-stationary solutions are perturbed or a pre-calculated model is
used, with similar concerns about the physical accuracy of the initial
conditions. In most simulations small parameter studies show that the
qualitative behaviour of the gravitational waves is robust; one
fundamental problem will be discussed later.

The key question for GWs and in particular for ET is whether numerical
relativity can compute sufficiently accurate signals. As this
crucially depends on the physics included in the model to be simulated
we will discuss the different input physics separately.

\subsection{Hydrodynamics}
\label{sec:matter_hydro}

Due to the scale and temperature of most interesting GW sources the
bulk of the matter can be reasonably modelled as a fluid, and in highly
dynamical situations such as a binary merger the bulk ideal fluid
effects, modelled by the relativistic Euler equations, will
dominate. These equations are typically written in conservation law
form~\cite{Banyuls97} as they admit discontinuous (weak) solutions for
which the conservation law form gives a unique entropy satisfying
solution. The microphysics under consideration is determined by the
EOS which closes the system of equations and is typically given as a
relation between the pressure and the independent thermodynamic and
composition variables.

The importance of the EOS is obvious, as has been illustrated by a
number of simulations (as reviewed by~\cite{0264-9381-26-11-114004},
but see in
particular~\cite{Oechslin:2007gn,Kiuchi:2009jt,Baiotti:2008ra}). Large
and small scale effects both in terms of the stiffness and the effects
of heat in simplified form have been covered. However current
uncertainty in the EOS is large, and at present the approaches
followed are to restrict to the ``best'' EOS currently available or to
use a simple parametrized EOS (\cite{Read:2008iy}) to best fit the
likely candidates. The effectiveness of the parametrized EOS combined
with numerical simulations has been illustrated by~\cite{Read:2009yp},
although it should be noted that the accuracy of the simulations
discussed there will not be sufficient for ET.

Numerical solutions of the relativistic Euler equations use techniques
from the standard Newtonian Euler equations. In particular it is
currently typical to use High Resolution Shock Capturing (HRSC)
methods (see e.g.~\cite{lrr-2008-7,lrr-2003-7} and references therein
for reviews). These ensure conservation is preserved (necessary to
ensure the correct weak solution is obtained if a shock is present)
and ensure that the solution converges near discontinuities. These
techniques are nonlinear in the sense that the information used at any
given point depends on the data in the neighbourhood of that point;
this is required to bypass the conditions of Godunov's theorem
(\cite{Godunov59}) and construct numerical methods that give correct
solutions that converge faster than first order.

The techniques currently used in simulations are extremely similar,
even when the spacetime is evolved using spectral methods
(\cite{Dimmelmeier:2004me,Duez:2008rb}), multiple patches
(\cite{Zink:2007xn}) or a mesh refined cartesian grid
(e.g.~\cite{Baiotti:2005vi,Yamamoto:2008js,Anderson:2008zp}). In
nearly every case a finite volume approach combined with a
reconstruction-evolution method and approximate Riemann solvers is
used, computing a conservative update over the neutron star and the
surrounding artificial atmosphere. These methods converge at best at
second order and are computationally expensive compared to the methods
required for the spacetime. A notable exception is the
smoothed-particle hydrodynamics simulations of
e.g.~\cite{Oechslin:2007gn,Oechslin02,Price:2006fi}, but as yet these
have not used full general relativity (but note the recent results
of~\cite{Rosswog:2009gn}).

These methods have been used to study purely hydrodynamical
scenarios. The numerical accuracy of the resulting GW signals is not
always easy to assess, with some work at lower resolutions not
claiming accuracy of better than 1 orbit (\cite{Read:2009yp}), whilst
others at the highest currently available resolutions claiming
accuracy in key quantities of the order of a percent
(\cite{Baiotti:2009gk}). Even in simulations with careful control of
the accuracy in the signal the amplitude accuracy is nearly
an order of magnitude worse than early black hole binary simulations
(compare figure 5 of \cite{Baiotti:2009gk} with figure 3 of
\cite{Baker:2006yw}). It is also clear that numerical accuracy is
significantly degraded when shocks appear such as in the complex
post-merger stage of a binary inspiral. Thus whilst it seems that
current techniques and computing power are sufficient to make
qualitative statements about gravitational waves and the general
effects of changes in the parameter space, detailed quantitative
predictions at the level required by ET are unlikely to result simply
by throwing computing power at current techniques, except in certain
simple cases.

Improvements in numerical techniques to significantly improve the
accuracy of the results, particularly in the complex post-merger
regime, are therefore needed to get the most out of numerical
relativity for ET. Extensions of standard finite volume techniques to
higher order (e.g.~\cite{Tchekhovskoy:2007zn}) are one obvious avenue
to explore. The implementation of more complex but flexible methods
such as Discontinuous Galerkin finite elements (as implemented in
relativity by \cite{Palenzuela:2008sf,Dumbser:2009hw}) is another
possibility. Finite difference methods have received some attention
(e.g.~\cite{Neilsen:2005rq,Anderson:2006ay}) and may be the simplest
approach. At present it seems unlikely that the highly accurate
spectral methods will provide a solution due to the problems these
methods encounter with discontinuities (although
\cite{Bonazzola:2006ii} presents one approach in a particular
approximation), but hybrid methods
(e.g.~\cite{Costa2007209,Costa2007970}) may allow for a combination of
the best features of spectral or simple high order schemes with HRSC
methods.

Finally it should be noted that the impact of the artificial
atmosphere used outside the compact GW source on the accuracy of the
results remains unclear, and methods to avoid the use of an atmosphere
(e.g.~\cite{Kastaun:2006ik,Duez:2002bn}) are at present not
sufficiently advantageous to use in relevant simulations. The presence
of a free surface (in the perfect fluid model) may be an impediment to
the use of more accurate numerical methods, and remains an area where
more understanding at both numerical and theoretical level is
needed. Purely on numerical grounds the use of finite elements is
attractive here, as the grid can be adapted to the surface of the
objects; all that is required is a consistent boundary condition to
impose.

\subsection{Magnetic fields}
\label{sec:matter_mhd}

So far the ideal MHD limit has been the focus of a range of studies
(for a recent review see~\cite{0264-9381-26-11-114004}), many simply
``adding on'' the magnetic field to study the qualitative differences
(see e.g.~\cite{Liu:2008xy,Giacomazzo:2009mp}), but some studying
additional instabilities that may arise
(e.g.~\cite{Price:2006fi,Kiuchi:2008ss}). With the increase in the
parameter space due to the strength and topology of the magnetic field
the choice of initial conditions becomes ever more important, and this
is where one fundamental difficulty remains.  It is expected that the
initial object will be approximately axi-stationary and the precise
topology and mixture of toroidal and poloidal components that would
make up this initial field has only been studied numerically under
certain assumptions,
see~\cite{Braithwaite06,Braithwaite08,Braithwaite09}. Until a
self-consistent quasi-stationary solution can be constructed the
standard method is to add a purely poloidal magnetic field to the
interior; it has been argued (\cite{Liu:2008xy}) that this is
sufficient for the inspiral phase of a magnetized binary.

In addition to the conservation constraints, numerical methods should
identically preserve the $\dB = 0$ constraint. There are many possible
approaches, with most codes implementing a variant of constrained
transport (e.g.~\cite{Toth2000}) and some using hyperbolic divergence
cleaning (e.g.~\cite{Anderson:2006ay}). Constrained transport methods
should ensure the constraint is satisfied to machine precision whilst
divergence cleaning propagates any errors away; however divergence
cleaning is simpler to extend to higher order methods and the presence
of the neutron star surface and artificial atmosphere may affect the
accuracy of constrained transport. In all cases the additional
constraints to be satisfied and fields to be evolved lead to lower
numerical accuracy; for example~\cite{Liu:2008xy} shows the numerical
error increasing by factors $\sim 5$ for evolutions of magnetic
binaries compared to non-magnetic NSs.

\subsection{Additional effects}
\label{sec:matter_additional}

In the context of computing GWs for ET the importance of physics
beyond the ideal MHD model are the changes to the internal structure,
which will certainly affect the fine detail and can change the
dynamics, leading to a qualitative change in the signal. The most
obvious problems arise through thermal effects (radiation transport
and diffusion post-merger or in core-collapse) or where the ideal
fluid model breaks down (viscous or multifluid effects such as
superfluidity, and in the crust).

\subsubsection{Radiation transport}
\label{sec:matter_radiation}

The transfer of heat and energy through radiation transport changes
the local structure of the fluid, and is known to be crucial in the
dynamics of core-collapse supernovae and may play a role in the
post-merger dynamics of a binary merger. However the simulation of the
full Boltzmann equation in $3+3+1$ dimensional general relativity
would seem to be impractical on timescales relevant for ET. An
estimate (\cite{Schnetter2007a}) suggests that even approximate
transport methods require peta scale computing power. A suggestion for
modelling the full problem is given by~\cite{Zink:2008qu}, but on the
timescale for ET it seems likely that approximation methods such
as~\cite{Farris:2008fe} will be the only practical solution.

\subsubsection{Non-ideal effects}
\label{sec:matter_nonideal}

The consideration of non-ideal effects may become important
post-merger when the matter surrounding the remnant can be both very
hot and yet not too dense, leading to a plasma where diffusion and
resistivity are important. Viscous effects have barely been touched on
-- for an isolated example see~\cite{Duez:2004nf}. Resistivity effects
have been considered and modelled in the magnetosphere, but modelling
for neutron stars is only just starting
(see~\cite{Palenzuela:2008sf,Dumbser:2009hw}).

The Newtonian limit gives mixed hyperbolic-parabolic systems of
equations with their associated causality problems. The relativistic
equivalents have stiff source terms leading to severe timestep
constraints when using standard numerical methods. Various numerical
schemes have been proposed to bypass this issue, including the use of
IMEX time integrators (\cite{Palenzuela:2008sf}). It seems plausible that
similar numerical techniques will be adopted as those for large mass
ratio binary black holes.

\subsubsection{Multifluid and elastic effects}
\label{sec:matter_super}

It is believed that most neutron stars have a region containing
superfluid neutron pairs and that magnetized neutron stars may be
superconducting in some region. The crust of a mature neutron star is
not a fluid but a lattice, best modelled as a relativistic elastic
solid inter-penetrated by an additional fluid. In addition, the
possible existence of exotic phases of matter in the inner core leads
to another region where multiple particles may be flowing freely and
independently. All of these effects involve the local interaction of
multiple fluids or solids (for a review see~\cite{Andersson:2006nr}),
with the appearance of additional dynamics and instabilities.

At present no numerical evolutions of any of these effects have been
performed. Frameworks for evolving multiple fluids have been worked
out (\cite{Andersson:2006nr}) but have not been tailored for numerical
work that would include shocks. It seems unlikely that multifluid
effects will be visible in numerical simulations in the inspiral
phase, but the additional propagation speeds and instabilities could
be visible near or post merger.

\section{Discussion: the nature of numerical relativity in the ET era}
\label{sec:discussion}

Over the last four decades, numerical relativists have been primarily 
concerned with the technical challenges of solving Einstein's equations
numerically: formulating Einstein's equations to be numerically stable,
developing appropriate gauge conditions, exploring the numerical
techniques necessary to accurately evolve relativistic spacetimes, and
to handle relativistic fluids and magnetic fields. Many of these issues
have been resolved to a sufficient degree that we can now extract
useful physics from the simulations. In addition, sensitive GW detectors are 
now in operation and the analysis of their data requires numerical
results. The field of numerical relativity is thus shifting from a focus on
theoretical and numerical-analysis issues to questions of astrophysics and GW data
analysis. 

Over the next decade, this shift will continue, and numerical
relativists will interact more closely with astronomers,
astrophysicists, and GW data analysts. Closer collaboration between
the NR and DA communities has already begun through the Numerical
{INJ}ection Analysis (NINJA) project, where numerical
black-hole-binary waveforms were injected into simulated Initial LIGO
and Virgo detector noise, and then the data were analyzed by a battery
of search and parameter estimation methods, to test their efficacy in
dealing with ``real'' GW signals. The successful first NINJA 
project~\cite{Aylott:2009ya} is now being followed with projects to more 
systematically test search pipelines. Among these are the suggestion to 
use actual detector data, which contain the full array of real detector artifacts
and can allow the most conclusive tests of search pipelines, and to extend
the original simulated-detector-noise NINJA strategy to matter sources 
(NS-NS, NS-BH and core collapse).
 
With the advent of Advanced LIGO/Virgo and the space-based detector
LISA~\cite{Danzmann:2003tv}, the interaction
between the NR and DA communities will surely become closer, and the
possibility of multi-messenger GW astronomy will include astronomers
and astrophysicists in these efforts, too. Once GW detections become
routine, and numerical simulations computationally cheaper and faster,
it may be typical for a candidate observation to motivate follow-up
numerical simulations, which in turn lead to more sophisticated data
analysis parameter estimation exercises, and comparison with
electromagnetic observations. The time scale of this back-and-forth
interaction may be as short as weeks or even days. For some regions of
the (at least black-hole-binary) parameter space, it is in fact likely
that GW astronomers will have ready access to numerical codes that
allow them to ``dial up'' any waveform they require for a particular
data-analysis exercise, and the focus of numerical-relativity research
will be only on those cases (for example, high mass ratios, binaries with 
complex spin interactions, and matter systems) that still present problems.

For matter simulations considerable work has been done to explore the
parameter space by including as much of the relevant physics as is
currently practical. However, in order to take full advantage of the
additional accuracy given by ET these simulations will have to improve
substantially in certain areas. Firstly the numerical methods
currently employed will probably not deliver the improvements in
accuracy required based on the (conservative) increase in computing
power projected here. Secondly the community will have to collaborate
closely with the data analysis community in order to determine which
aspects of the physical model will actually lead to interesting
observable effects, and hence where in the parameter space the
numerical simulations should be focused. As the discussion in
section~\ref{sec:matter} makes clear it would be possible to spend all
the time and increased computing power in applying and simulating
better physical models and exploring the parameter space. It seems
likely that to take the greatest advantage of ET a narrower focus is
needed to produce the most accurate simulations including the most
relevant physics for the best candidate sources.

\begin{acknowledgements}
  MH was supported by SFI grant 07/RFP/PHYF148, and thanks the
  University of the Balearic Islands for hospitality while some of
  this work was completed, and Sascha Husa for numerous
  discussions. IH was supported by the STFC rolling grant PP/E001025/1
  and Nuffield Foundation grant NAL/32622.
\end{acknowledgements}

\bibliographystyle{spphys}       
\bibliography{NRforET}   

\begin{thebibliography}{100}
\providecommand{\url}[1]{{#1}}
\providecommand{\urlprefix}{URL }
\expandafter\ifx\csname urlstyle\endcsname\relax
  \providecommand{\doi}[1]{DOI \discretionary{}{}{}#1}\else
  \providecommand{\doi}{DOI \discretionary{}{}{}\begingroup
  \urlstyle{rm}\Url}\fi

\bibitem{Abbott:2007kv}
B.~Abbott, et~al.,   (2007).
\newblock {arXiv:0711.3041}

\bibitem{Acernese2006}
F.~Acernese, et~al., Class. Quantum Grav. \textbf{23}, S635 (2006)

\bibitem{GEOStatus:2006}
S.~{Hild (for the LIGO Scientific Collaboration)}, Class. Quantum Grav.
  \textbf{23}, S643 (2006)

\bibitem{Sathyaprakash:2009xs}
B.S. Sathyaprakash, B.F. Schutz, Living Rev. Rel. \textbf{12}, 2 (2009)

\bibitem{Pretorius:2005gq}
F.~Pretorius, Phys. Rev. Lett. \textbf{95}, 121101 (2005).
\newblock \doi{10.1103/PhysRevLett.95.121101}

\bibitem{Campanelli:2005dd}
M.~Campanelli, C.O. Lousto, P.~Marronetti, Y.~Zlochower, Phys. Rev. Lett.
  \textbf{96}, 111101 (2006)

\bibitem{Baker:2005vv}
J.G. Baker, J.~Centrella, D.I. Choi, M.~Koppitz, J.~van Meter, Phys. Rev. Lett.
  \textbf{96}, 111102 (2006)

\bibitem{Hannam:2009rd}
M.~Hannam, Class. Quant. Grav. \textbf{26}, 114001 (2009).
\newblock \doi{10.1088/0264-9381/26/11/114001}

\bibitem{Shibata:1999wm}
M.~Shibata, K.~Uryu, Phys. Rev. \textbf{D61}, 064001 (2000).
\newblock \doi{10.1103/PhysRevD.61.064001}

\bibitem{0264-9381-26-11-114004}
J.~Faber, Classical and Quantum Gravity \textbf{26}(11), 114004 (18pp) (2009).
\newblock \urlprefix\url{http://stacks.iop.org/0264-9381/26/114004}

\bibitem{Pfeiffer:2007yz}
H.P. Pfeiffer, et~al., Class. Quant. Grav. \textbf{24}, S59 (2007).
\newblock \doi{10.1088/0264-9381/24/12/S06}

\bibitem{Husa:2007rh}
S.~Husa, M.~Hannam, J.A. Gonzalez, U.~Sperhake, B.~Br{\"u}gmann, Phys. Rev.
  \textbf{D77}, 044037 (2008).
\newblock \doi{10.1103/PhysRevD.77.044037}

\bibitem{Campanelli:2008nk}
M.~Campanelli, C.O. Lousto, H.~Nakano, Y.~Zlochower, Phys. Rev. \textbf{D79},
  084010 (2009)

\bibitem{Alcubierre2008}
M.~Alcubierre, \emph{Introduction to 3+1 Numerical Relativity} (Oxford
  University Press, USA, 2008)

\bibitem{Cook:2000vr}
G.B. Cook, Living Rev. Rel. \textbf{3}, 5 (2000)

\bibitem{Pretorius:2007nq}
F.~Pretorius, in \emph{Physics of relativistic objects in compact binaries:
  from birth to coalescence}, vol. 359, ed. by M.~Colpi, P.~Casella, V.~Gorini,
  U.~Moschella, A.~Possenti (Springer-Verlag, 2009), vol. 359, pp. 269--305.
\newblock {arXiv:0710.1338}

\bibitem{Husa:2008jx}
S.~Husa, Eur. Phys. J. ST \textbf{152}, 183 (2007)

\bibitem{Palenzuela:2009yr}
C.~Palenzuela, M.~Anderson, L.~Lehner, S.L. Liebling, D.~Neilsen,   (2009).
\newblock {arXiv:0905.1121}

\bibitem{vanMeter:2009gu}
J.R. van Meter, et~al.,   (2009).
\newblock {arXiv:0908.0023}

\bibitem{Lindblom:2008cm}
L.~Lindblom, B.J. Owen, D.A. Brown, Phys. Rev. \textbf{D78}, 124020 (2008).
\newblock \doi{10.1103/PhysRevD.78.124020}

\bibitem{Hannam:2009hh}
M.~Hannam, et~al., Phys. Rev. \textbf{D79}, 084025 (2009)

\bibitem{Reisswig:2009us}
C.~Reisswig, N.T. Bishop, D.~Pollney, B.~Szilagyi,   (2009).
\newblock {arXiv:0907.2637}

\bibitem{Scheel:2008rj}
M.A. Scheel, et~al., Phys. Rev. \textbf{D79}, 024003 (2009).
\newblock \doi{10.1103/PhysRevD.79.024003}

\bibitem{Sintes:1999cg}
A.M. Sintes, A.~Vecchio,   (1999).
\newblock {gr-qc/0005058}

\bibitem{VanDenBroeck:2006ar}
C.~Van Den~Broeck, A.S. Sengupta, Class. Quant. Grav. \textbf{24}, 1089 (2007).
\newblock \doi{10.1088/0264-9381/24/5/005}

\bibitem{Babak:2008bu}
S.~Babak, M.~Hannam, S.~Husa, B.~Schutz,   (2008).
\newblock {arXiv:0806.1591}

\bibitem{Thorpe:2008wh}
J.I. Thorpe, et~al.,   (2008).
\newblock {arXiv:0811.0833}

\bibitem{Boyle:2008ge}
M.~Boyle, et~al., Phys. Rev. \textbf{D78}, 104020 (2008).
\newblock \doi{10.1103/PhysRevD.78.104020}

\bibitem{PollneyNRDA}
D.~Pollney, Talk given at NRDA2009, AEI-Potsdam, July 6-9, 2009

\bibitem{Buonanno:1998gg}
A.~Buonanno, T.~Damour, Phys. Rev. D \textbf{59}, 084006 (1999)

\bibitem{Buonanno00a}
A.~Buonanno, T.~Damour, Phys. Rev. D \textbf{62}, 064015 (2000)

\bibitem{Damour:2000we}
T.~Damour, P.~Jaranowski, G.~Sch{\"a}fer, Phys. Rev. D \textbf{62}, 084011
  (2000)

\bibitem{Damour:2001tu}
T.~Damour, Phys. Rev. D \textbf{64}, 124013 (2001)

\bibitem{Buonanno:2005xu}
A.~Buonanno, Y.~Chen, T.~Damour, Phys. Rev. \textbf{D74}, 104005 (2006)

\bibitem{Buonanno:2007pf}
A.~Buonanno, et~al., Phys. Rev. \textbf{D76}, 104049 (2007)

\bibitem{Damour:2007yf}
T.~Damour, A.~Nagar, Phys. Rev. \textbf{D77}, 024043 (2008).
\newblock \doi{10.1103/PhysRevD.77.024043}

\bibitem{Damour:2007vq}
T.~Damour, A.~Nagar, E.N. Dorband, D.~Pollney, L.~Rezzolla, Phys. Rev.
  \textbf{D77}, 084017 (2008).
\newblock \doi{10.1103/PhysRevD.77.084017}

\bibitem{Damour:2008te}
T.~Damour, A.~Nagar, M.~Hannam, S.~Husa, B.~Br{\"u}gmann, Phys. Rev.
  \textbf{D78}, 044039 (2008).
\newblock \doi{10.1103/PhysRevD.78.044039}

\bibitem{Baker:2008mj}
J.G. Baker, et~al., Phys. Rev. \textbf{D78}, 044046 (2008).
\newblock \doi{10.1103/PhysRevD.78.044046}

\bibitem{Mroue:2008fu}
A.H. Mroue, L.E. Kidder, S.A. Teukolsky, Phys. Rev. \textbf{D78}, 044004
  (2008).
\newblock \doi{10.1103/PhysRevD.78.044004}

\bibitem{Damour:2009kr}
T.~Damour, A.~Nagar,   (2009).
\newblock {arXiv:0902.0136}

\bibitem{Buonanno:2009qa}
A.~Buonanno, et~al.,   (2009).
\newblock {arXiv:0902.0790}

\bibitem{Buonanno:2006ui}
A.~Buonanno, G.B. Cook, F.~Pretorius, Phys. Rev. \textbf{D75}, 124018 (2007).
\newblock \doi{10.1103/PhysRevD.75.124018}

\bibitem{Ajith:2007qp}
P.~Ajith, et~al., Class. Quant. Grav. \textbf{24}, S689 (2007).
\newblock \doi{10.1088/0264-9381/24/19/S31}

\bibitem{Ajith:2007kx}
P.~Ajith, et~al., Phys. Rev. \textbf{D77}, 104017 (2008).
\newblock \doi{10.1103/PhysRevD.77.104017}

\bibitem{Ajith:2007xh}
P.~Ajith, Class. Quant. Grav. \textbf{25}, 114033 (2008).
\newblock \doi{10.1088/0264-9381/25/11/114033}

\bibitem{Ajith:prep}
P.~Ajith, et~al.,   (2009).
\newblock In preparation

\bibitem{Shibata:2009cn}
M.~Shibata, K.~Kyutoku, T.~Yamamoto, K.~Taniguchi, Phys. Rev. \textbf{D79},
  044030 (2009).
\newblock \doi{10.1103/PhysRevD.79.044030}

\bibitem{Baker:2006ha}
J.G. Baker, J.R. van Meter, S.T. McWilliams, J.~Centrella, B.J. Kelly, Phys.
  Rev. Lett. \textbf{99}, 181101 (2007).
\newblock \doi{10.1103/PhysRevLett.99.181101}

\bibitem{Hannam:2007ik}
M.~Hannam, S.~Husa, U.~Sperhake, B.~Br{\"u}gmann, J.A. Gonzalez, Phys. Rev.
  \textbf{D77}, 044020 (2008).
\newblock \doi{10.1103/PhysRevD.77.044020}

\bibitem{Gopakumar:2007vh}
A.~Gopakumar, M.~Hannam, S.~Husa, B.~Bruegmann, Phys. Rev. \textbf{D78}, 064026
  (2008).
\newblock \doi{10.1103/PhysRevD.78.064026}

\bibitem{Boyle:2007ft}
M.~Boyle, et~al., Phys. Rev. \textbf{D76}, 124038 (2007).
\newblock \doi{10.1103/PhysRevD.76.124038}

\bibitem{Hannam:2007wf}
M.~Hannam, S.~Husa, B.~Bruegmann, A.~Gopakumar, Phys. Rev. \textbf{D78}, 104007
  (2008).
\newblock \doi{10.1103/PhysRevD.78.104007}

\bibitem{Hinder:2008kv}
I.~Hinder, F.~Herrmann, P.~Laguna, D.~Shoemaker,   (2008)

\bibitem{Hannam:prep}
M.~Hannam, et~al.,   (2009).
\newblock In preparation

\bibitem{Buonanno:2009zt}
A.~Buonanno, B.~Iyer, E.~Ochsner, Y.~Pan, B.S. Sathyaprakash,   (2009)

\bibitem{Boyle:2007sz}
L.~Boyle, M.~Kesden, S.~Nissanke, Phys. Rev. Lett. \textbf{100}, 151101 (2008).
\newblock \doi{10.1103/PhysRevLett.100.151101}

\bibitem{Boyle:2007ru}
L.~Boyle, M.~Kesden, Phys. Rev. \textbf{D78}, 024017 (2008).
\newblock \doi{10.1103/PhysRevD.78.024017}

\bibitem{Volonteri:2004cf}
M.~Volonteri, P.~Madau, E.~Quataert, M.J. Rees, Astrophys. J. \textbf{620}, 69
  (2005).
\newblock \doi{10.1086/426858}

\bibitem{Gammie:2003qi}
C.F. Gammie, S.L. Shapiro, J.C. McKinney, Astrophys. J. \textbf{602}, 312
  (2004)

\bibitem{Shapiro05}
S.L. {Shapiro}, Astrophys. J. \textbf{620}, 59 (2005)

\bibitem{Lovelace:2008tw}
G.~Lovelace, R.~Owen, H.P. Pfeiffer, T.~Chu, Phys. Rev. \textbf{D78}, 084017
  (2008).
\newblock \doi{10.1103/PhysRevD.78.084017}

\bibitem{Lovelace:2008hd}
G.~Lovelace, Class. Quant. Grav. \textbf{26}, 114002 (2009).
\newblock \doi{10.1088/0264-9381/26/11/114002}

\bibitem{Hannam:2006zt}
M.~Hannam, S.~Husa, B.~Br{\"u}gmann, J.A. Gonzalez, U.~Sperhake, Class. Quantum
  Grav. \textbf{24}, S15 (2007)

\bibitem{Gonzalez:2008bi}
J.A. Gonzalez, U.~Sperhake, B.~Brugmann, Phys. Rev. \textbf{D79}, 124006
  (2009).
\newblock \doi{10.1103/PhysRevD.79.124006}

\bibitem{Husa:2007hp}
S.~Husa, J.A. Gonzalez, M.~Hannam, B.~Br{\"u}gmann, U.~Sperhake, Class. Quant.
  Grav. \textbf{25}, 105006 (2008).
\newblock \doi{10.1088/0264-9381/25/10/105006}

\bibitem{Lousto:2007rj}
C.O. Lousto, Y.~Zlochower, Phys. Rev. \textbf{D77}, 024034 (2008).
\newblock \doi{10.1103/PhysRevD.77.024034}

\bibitem{Pazos:2009vb}
E.~Pazos, M.~Tiglio, M.D. Duez, L.E. Kidder, S.A. Teukolsky,   (2009).
\newblock {arXiv:0904.0493}

\bibitem{lrr-2009-3}
J.~Winicour, Living Reviews in Relativity \textbf{12}(3) (2009).
\newblock \urlprefix\url{http://www.livingreviews.org/lrr-2009-3}

\bibitem{Pretorius:2004jg}
F.~Pretorius, Class. Quantum Grav. \textbf{22}, 425 (2005)

\bibitem{Pretorius:2006tp}
F.~Pretorius, Class. Quantum Grav. \textbf{23}, S529 (2006)

\bibitem{Lindblom:2005qh}
L.~Lindblom, M.A. Scheel, L.E. Kidder, R.~Owen, O.~Rinne, Class. Quantum Grav.
  \textbf{23}, S447 (2006)

\bibitem{Lau:2008fb}
S.R. Lau, H.P. Pfeiffer, J.S. Hesthaven,   (2008).
\newblock {arXiv:0808.2597}

\bibitem{Brown:2005jc}
J.D. Brown, Phys. Rev. \textbf{D73}, 024001 (2006).
\newblock \doi{10.1103/PhysRevD.73.024001}

\bibitem{Richter:2008pr}
R.~Richter, C.~Lubich, Class. Quant. Grav. \textbf{25}, 225018 (2008).
\newblock \doi{10.1088/0264-9381/25/22/225018}

\bibitem{Richter:2009ff}
R.~Richter, Class. Quant. Grav. \textbf{26}, 145017 (2009).
\newblock \doi{10.1088/0264-9381/26/14/145017}

\bibitem{Frauendiener:2008bt}
J.~Frauendiener,   (2008).
\newblock {arXiv:0805.4465}

\bibitem{Ohme:2009gn}
F.~Ohme, M.~Hannam, S.~Husa, N.O. Murchadha,   (2009).
\newblock {arXiv:0905.0450}

\bibitem{Buchman:2009ew}
L.T. Buchman, H.P. Pfeiffer, J.M. Bardeen,   (2009).
\newblock {arXiv:0907.3163}

\bibitem{Moncrief:2008ie}
V.~Moncrief, O.~Rinne, Class. Quant. Grav. \textbf{26}, 125010 (2009).
\newblock \doi{10.1088/0264-9381/26/12/125010}

\bibitem{Hannam:2006vv}
M.~Hannam, S.~Husa, D.~Pollney, B.~Bruegmann, N.~O'Murchadha, Phys. Rev. Lett.
  \textbf{99}, 241102 (2007).
\newblock \doi{10.1103/PhysRevLett.99.241102}

\bibitem{Hannam:2008sg}
M.~Hannam, S.~Husa, F.~Ohme, B.~Bruegmann, N.~O'Murchadha, Phys. Rev.
  \textbf{D78}, 064020 (2008).
\newblock \doi{10.1103/PhysRevD.78.064020}

\bibitem{Hannam:2009ib}
M.~Hannam, S.~Husa, N.O. Murchadha,   (2009).
\newblock {arXiv:0908.1063}

\bibitem{Ott:2008wt}
C.D. Ott, Class. Quant. Grav. \textbf{26}, 063001 (2009).
\newblock \doi{10.1088/0264-9381/26/6/063001}

\bibitem{Oechslin:2007gn}
R.~Oechslin, H.T. Janka, Phys. Rev. Lett. \textbf{99}, 121102 (2007).
\newblock \doi{10.1103/PhysRevLett.99.121102}

\bibitem{Barnes:2004if}
A.P. Barnes, P.G. Lefloch, B.G. Schmidt, J.M. Stewart, Class. Quant. Grav.
  \textbf{21}, 5043 (2004).
\newblock \doi{10.1088/0264-9381/21/22/003}

\bibitem{Banyuls97}
F.~Banyuls, J.A. Font, J.M. Ib{\'a}{\~n}ez, J.M. Mart{\'\i}, J.A. Miralles,
  Astrophys. J. \textbf{476}, 221 (1997)

\bibitem{Kiuchi:2009jt}
K.~Kiuchi, Y.~Sekiguchi, M.~Shibata, K.~Taniguchi,   (2009).
\newblock {arXiv:0904.4551}

\bibitem{Baiotti:2008ra}
L.~Baiotti, B.~Giacomazzo, L.~Rezzolla, Phys. Rev. \textbf{D78}, 084033 (2008).
\newblock \doi{10.1103/PhysRevD.78.084033}

\bibitem{Read:2008iy}
J.S. Read, B.D. Lackey, B.J. Owen, J.L. Friedman, Phys. Rev. \textbf{D79},
  124032 (2009).
\newblock \doi{10.1103/PhysRevD.79.124032}

\bibitem{Read:2009yp}
J.S. Read, et~al., Phys. Rev. \textbf{D79}, 124033 (2009).
\newblock \doi{10.1103/PhysRevD.79.124033}

\bibitem{lrr-2008-7}
J.A. Font, Living Reviews in Relativity \textbf{11}(7) (2008).
\newblock \urlprefix\url{http://www.livingreviews.org/lrr-2008-7}

\bibitem{lrr-2003-7}
J.M. Mart\'i, E.~M\"uller, Living Reviews in Relativity \textbf{6}(7) (2003).
\newblock \urlprefix\url{http://www.livingreviews.org/lrr-2003-7}

\bibitem{Godunov59}
S.K. Godunov, Mat. Sb. \textbf{47}, 271 (1959)

\bibitem{Dimmelmeier:2004me}
H.~Dimmelmeier, J.~Novak, J.A. Font, J.M. Ibanez, E.~Muller, Phys. Rev.
  \textbf{D71}, 064023 (2005).
\newblock \doi{10.1103/PhysRevD.71.064023}

\bibitem{Duez:2008rb}
M.D. Duez, et~al., Phys. Rev. \textbf{D78}, 104015 (2008).
\newblock \doi{10.1103/PhysRevD.78.104015}

\bibitem{Zink:2007xn}
B.~Zink, E.~Schnetter, M.~Tiglio, Phys. Rev. \textbf{D77}, 103015 (2008).
\newblock \doi{10.1103/PhysRevD.77.103015}

\bibitem{Baiotti:2005vi}
L.~Baiotti, I.~Hawke, L.~Rezzolla, E.~Schnetter, Phys. Rev. Lett. \textbf{94},
  131101 (2005).
\newblock \doi{10.1103/PhysRevLett.94.131101}

\bibitem{Yamamoto:2008js}
T.~Yamamoto, M.~Shibata, K.~Taniguchi, Phys. Rev. \textbf{D78}, 064054 (2008).
\newblock \doi{10.1103/PhysRevD.78.064054}

\bibitem{Anderson:2008zp}
M.~Anderson, et~al., Phys. Rev. Lett. \textbf{100}, 191101 (2008).
\newblock \doi{10.1103/PhysRevLett.100.191101}

\bibitem{Oechslin02}
R.~Oechslin, S.~Rosswog, F.~Thielemann, Phys.Rev. D \textbf{65}, 103005 (2002)

\bibitem{Price:2006fi}
D.J. Price, S.~Rosswog, Science \textbf{312}(5774), 719 (2006).
\newblock \doi{10.1126/science.1125201}

\bibitem{Rosswog:2009gn}
S.~Rosswog,   (2009).
\newblock {arXiv:0907.4890}

\bibitem{Baiotti:2009gk}
L.~Baiotti, B.~Giacomazzo, L.~Rezzolla, Class. Quant. Grav. \textbf{26}, 114005
  (2009).
\newblock \doi{10.1088/0264-9381/26/11/114005}

\bibitem{Baker:2006yw}
J.G. Baker, J.~Centrella, D.I. Choi, M.~Koppitz, J.~van Meter, Phys. Rev.
  \textbf{D73}, 104002 (2006).
\newblock \doi{10.1103/PhysRevD.73.104002}

\bibitem{Tchekhovskoy:2007zn}
A.~Tchekhovskoy, J.C. McKinney, R.~Narayan, Mon. Not. Roy. Astron. Soc.
  \textbf{379}, 469 (2007).
\newblock \doi{10.1111/j.1365-2966.2007.11876.x}

\bibitem{Palenzuela:2008sf}
C.~Palenzuela, L.~Lehner, O.~Reula, L.~Rezzolla,   (2008).
\newblock {arXiv:0810.1838}

\bibitem{Dumbser:2009hw}
M.~Dumbser, O.~Zanotti,   (2009).
\newblock {arXiv:0903.4832}

\bibitem{Neilsen:2005rq}
D.~Neilsen, E.W. Hirschmann, R.S. Millward, Class. Quant. Grav. \textbf{23},
  S505 (2006).
\newblock \doi{10.1088/0264-9381/23/16/S12}

\bibitem{Anderson:2006ay}
M.~Anderson, E.~Hirschmann, S.L. Liebling, D.~Neilsen, Class. Quant. Grav.
  \textbf{23}, 6503 (2006).
\newblock \doi{10.1088/0264-9381/23/22/025}

\bibitem{Bonazzola:2006ii}
S.~Bonazzola, L.~Villain, M.~Bejger, Class. Quant. Grav. \textbf{24}, S221
  (2007).
\newblock \doi{10.1088/0264-9381/24/12/S15}

\bibitem{Costa2007209}
B.~Costa, W.S. Don, Journal of Computational and Applied Mathematics
  \textbf{204}(2), 209  (2007).
\newblock \doi{DOI: 10.1016/j.cam.2006.01.039}

\bibitem{Costa2007970}
B.~Costa, W.S. Don, Journal of Computational Physics \textbf{224}(2), 970
  (2007).
\newblock \doi{DOI: 10.1016/j.jcp.2006.11.002}

\bibitem{Kastaun:2006ik}
W.~Kastaun, Phys. Rev. \textbf{D74}, 124024 (2006).
\newblock \doi{10.1103/PhysRevD.74.124024}

\bibitem{Duez:2002bn}
M.D. Duez, P.~Marronetti, S.L. Shapiro, T.W. Baumgarte, Phys. Rev.
  \textbf{D67}, 024004 (2003).
\newblock \doi{10.1103/PhysRevD.67.024004}

\bibitem{Liu:2008xy}
Y.T. Liu, S.L. Shapiro, Z.B. Etienne, K.~Taniguchi, Phys. Rev. \textbf{D78},
  024012 (2008).
\newblock \doi{10.1103/PhysRevD.78.024012}

\bibitem{Giacomazzo:2009mp}
B.~Giacomazzo, L.~Rezzolla, L.~Baiotti,   (2009).
\newblock {arXiv:0901.2722}

\bibitem{Kiuchi:2008ss}
K.~Kiuchi, M.~Shibata, S.~Yoshida, Phys. Rev. \textbf{D78}, 024029 (2008).
\newblock \doi{10.1103/PhysRevD.78.024029}

\bibitem{Braithwaite06}
J.~Braithwaite, A.~Nordlund, A\&A \textbf{450}(3), 1077 (2006).
\newblock \doi{10.1051/0004-6361:20041980}.
\newblock \urlprefix\url{http://dx.doi.org/10.1051/0004-6361:20041980}

\bibitem{Braithwaite08}
J.~Braithwaite, MNRAS \textbf{386}, 1947 (2008)

\bibitem{Braithwaite09}
J.~Braithwaite, MNRAS \textbf{397}, 763 (2009)

\bibitem{Toth2000}
G.~Toth, J. Comput. Phys. \textbf{161}, 605 (2000)

\bibitem{Schnetter2007a}
E.~Schnetter, C.D. Ott, G.~Allen, P.~Diener, T.~Goodale, T.~Radke, E.~Seidel,
  J.~Shalf, in \emph{Petascale Computing: Algorithms and Applications}, ed. by
  D.A. Bader (Chapman \& Hall/CRC Computational Science Series, 2007), chap.~24

\bibitem{Zink:2008qu}
B.~Zink,   (2008).
\newblock {arXiv:0810.5349}

\bibitem{Farris:2008fe}
B.D. Farris, T.K. Li, Y.T. Liu, S.L. Shapiro, Phys. Rev. \textbf{D78}, 024023
  (2008).
\newblock \doi{10.1103/PhysRevD.78.024023}

\bibitem{Duez:2004nf}
M.D. Duez, Y.T. Liu, S.L. Shapiro, B.C. Stephens, Phys. Rev. \textbf{D69},
  104030 (2004).
\newblock \doi{10.1103/PhysRevD.69.104030}

\bibitem{Andersson:2006nr}
N.~Andersson, G.L. Comer, Living Rev. Rel. \textbf{10}, 1 (2005)

\bibitem{Aylott:2009ya}
B.~Aylott, et~al., Class. Quant. Grav. \textbf{26}, 165008 (2009).
\newblock \doi{10.1088/0264-9381/26/16/165008}

\bibitem{Danzmann:2003tv}
K.~Danzmann, A.~R{\"u}diger, Classical Quantum Gravity \textbf{20}, S1 (2003).
\newblock \urlprefix\url{stacks.iop.org/CQG/20/S2}

\end{thebibliography}

\end{document}